\documentclass[sigconf]{acmart} 

\AtBeginDocument{%
  \providecommand\BibTeX{{%
    \normalfont B\kern-0.5em{\scshape i\kern-0.25em b}\kern-0.8em\TeX}}}

\setcopyright{acmcopyright}
\copyrightyear{2018}
\acmYear{2018}
\acmDOI{XXXXXXX.XXXXXXX}

\acmConference[Conference acronym 'XX]{Make sure to enter the correct
  conference title from your rights confirmation email}{June 03--05,
  2018}{Woodstock, NY}
\acmPrice{15.00}
\acmISBN{978-1-4503-XXXX-X/18/06}

\usepackage[whole]{bxcjkjatype} 
\usepackage{enumerate} 

\usepackage{caption}
\usepackage{subcaption}

\def\tahera{\tahera}
\def\tahera{\textcolor{red}}
\begin{document}

\title{Toward Pioneering Sensors and Features \\ Using Large Language Models in Human Activity Recognition}


\author{HARU KANEKO}
\orcid{0000-0002-1881-5182}
\affiliation{%
  \institution{Kyushu Institute of Technology}
  \streetaddress{2-4, Hibikino,Wakamatsu ward}
  \city{Kitakyushu}
  \state{Fukuoka}
  \country{Japan}}
\email{kaneko.haru193@mail.kyutech.jp}

\author{SOZO INOUE}
\orcid{0000-0003-1109-8130}
\affiliation{%
  \institution{Kyushu Institute of Technology}
  \streetaddress{2-4, Hibikino,Wakamatsu ward}
  \city{Kitakyushu}
  \state{Fukuoka}
  \country{Japan}
}
\email{sozo@brain.kyutech.ac.jp}

\begin{abstract}
In this paper, we propose a feature pioneering method using Large Language Models (LLMs). 
In the proposed method, we use ChatGPT~\footnote{https://chat.openai.com/chatm} to find new sensor locations and new features. 
Then we evaluate the machine learning model which uses the found features using Opportunity Dataset ~\cite{OPPortunityDataset1, OPPortunityDataset2}. 
In current machine learning, humans make features, for this engineers visit real sites and have discussions with experts and veteran workers. 
However, this method has the problem that the quality of the features depends on the engineer. 
In order to solve this problem, we propose a way to make new features using LLMs. 
As a result, we obtain almost the same level of accuracy as the proposed model which used fewer sensors and the model uses all sensors in the dataset. 
This indicates that the proposed method is able to extract important features efficiently. 
\end{abstract}






\maketitle
\section{Introduction}
\label{section:introduction}
Human Activity Recognition (HAR) using wearables refers to the process of automatically identifying and classifying human activities based on data collected from wearable devices. 
These wearables, which include smartwatches, fitness trackers, and motion sensors, are outfitted with sensors such as accelerometers, gyroscopes, and heart rate monitors~\cite{lara2012survey, mar2019}. 
These sensors record information about the user's movements, body positions, and physiological responses. HAR algorithms analyze this data to recognize and categorize activities such as walking, running, cycling, sleeping, and others. 
HAR systems utilize machine learning techniques to learn patterns, create accurate real-time models, and differentiate between various activities by selecting informative and discriminative features. 

However, developing features from raw data requires both domain knowledge and human efforts. 
Later, choosing features for the best fit can take time as well. 
Despite these efforts, contemporary machine learning methods predominantly rely on features manually crafted by humans to determine which values are pertinent for estimation. 
In particular, to reflect the field/domain-specific knowledge in the features, the people visited the field where the data was from and had discussions with specialists/veteran workers in that field, when they made features. 
However, those reflections of domain knowledge by such methods depend on the experience and knowledge of the data scientists who create features.

In this paper, we propose a feature pioneering method using Large Language Models (LLMs). 
We input ChatGPT about the machine learning task which is we want to do and the current sensor location and ask ChatGPT where we should install new sensors in the human body.
Then based on the output of ChatGPT, we add new sensors and new features.

In the experiment, we assume that only elbow sensor data are used for activity recognition, then ask the location of sensors that should be added to ChatGPT.
As a result, we got 7 new sensor locations and 14 new features suggested by ChatGPT. 
Then we evaluate the human activity recognition model using Opportunity Dataset ~\cite{OPPortunityDataset1, OPPortunityDataset2} which mode of locomotion. 
In the result of HAR models, we obtain almost the same level of accuracy as the proposed model which used fewer sensors and the model uses all sensors in the dataset. 
This indicates that the proposed method is able to extract important features efficiently. 



\section{Background}
\label{section:background}


Basically, machine learning uses features created by humans to figure out what values are useful for estimation. 
Additionally, to reflect the field/domain-specific knowledge in the features, people visited the field where the data was from, and had discussions with specialists/veteran workers in that field, when they make features. 
However, those reflections of domain knowledge by such methods rely on the experience and knowledge of the data scientists who create features. 


In the machine learning area, also in the human activity recognition area, there are methods to evaluate 
the effectiveness of sensors in prediction models, there are methodologies available in the HAR field. Examining feature importance is one method of doing this~\cite{featureImportance, featureImportance2} or SHAP value(SHapley Additive exPlanation) ~\cite{featureShapValue, featureShapValue2}. 
Utilizing a technique called "feature importance evaluation",  researchers can determine the importance and relevance of specific features or sensors in terms of how much they contribute to a model's accuracy and prognostication. 
This evaluation process helps determine how well sensors capture pertinent data for activity recognition by quantifying the influence and impact of various features. 
It enables the creation of robust prediction models by enabling researchers to prioritize and pick the most important features or sensors. 
The machine learning community can choose sensors and create features with greater knowledge by using feature importance evaluation, which will ultimately improve performance. 
However, these methods can only use after data collection. 

Recently, Large language models (LLMs) like GPT-4, which are incredibly powerful, are now being used in a variety of applications, including productivity suites, code generation tools, and search engines. 
These generative models are anticipated to have an impact across industries, altering how we approach tasks like writing, programming, design, and more ~\cite{agrawal2022chatgpt, will}. 
Such, prompt engineering has been attracting a lot of attention. 
This is a technique that optimizes the input text and improves the quality of the output when using text input as the UI for a language model called LLM. 
Due to the increasing performance of LLMs, many prompt patterns have been developed ~\cite{promptPattern, promptPattern2}.

For this reason, in this paper, we aim to solve the problem that the features making are depending on the skill of engineers, and we propose prompts for pioneering features. 
For this, we make a prompt to find new sensor locations and new features for human activity recognition using ChatGPT. 
Then, assuming 2 cases the case before data collection and the case after data collection, we propose a feature pioneering method for both cases. 
In the case of before data collection, the proposed prompts use only information from the task of activity recognition, and in the case of after data collection, the proposed prompt also uses additional information from the result of the human activity recognition model. 

\section{Method}
\label{section:proposedMethod}
In this section, we propose a feature pioneering prompt for LLMs and evaluate the machine learning model which uses features funded by the proposed method. 
For this evaluation, we use the Opportunity Dataset which is human activities of daily living (ADL) dataset including activity labels and IMU/accelerometer sensor data.

\subsection{Problem Setup} 
\label{section:probremSetting}
The problem setup is the case where data collection and machine learning have already been done, then want to further data collection to improve the accuracy. 
In this case, we propose the sensor pioneering method (in section ~\ref{section:sensorExpansion}) and the feature augmentation method (in section ~\ref{section:featureCalculation}). 

In the sensor pioneering method, we propose to find new sensor locations that are the locations of human surfaces to be useful for recognizing human activity. 
In the feature augmentation method, we propose to find new feature calculations such as mean, variance, etc. for recognizing human activity. 

\subsection{Sensor Pioneering}

\label{section:sensorExpansion}

Our prompt is made up of the following 5 structures, (1) ``Your role'', (2) ``The problem you need to solve'', (3) ``Output activity labels'', (4) ``Current features'', (5) ``Current result'', and (6) ``Your task''.


\subsubsection*{\textbf{(1) ``Your role'' :}} 
Here defines the position of the skilled person from whom you wish to obtain domain knowledge.

\subsubsection*{\textbf{(2) ``The problem you need to solve'' :}} 
Here explain what kind of machine learning task.
Explain what you want to do, what the task is, and what data is available.
At this point, use specialized words such as IMU and acc.

\subsubsection*{\textbf{(3) ``Output activity labels'' :}} 
Here we show the prediction labels. 
In this case, the labels are shown as words rather than IDs, etc., so that the meaning of the labels can be understood.

\subsubsection*{\textbf{(4) ``Current features'' :}} 
Here we explain the current location of the sensor. 
At this point, explain using the words of position in the body.

\subsubsection*{\textbf{(5) ``Current result'' :}} 
Here we describe the results of the current prediction model. 
In particular, in order to obtain features which is effective for ignoring misclassification, 
we describe the case of often misses classification based on the confusion matrix. 

\subsubsection*{\textbf{(6) ``Your task'' :}} 
Here we indicate what you want ChatGPT to output.
Since the purpose of this paper is to extend features, we specify that new features should be output.


\subsection{Feature Augmentation}
\label{section:featureCalculation}
This prompt is made up of 4 structures, and the following 2 structures (1) ``The problem you need to solve'', and (2) ``Output activity labels'' are the same with section ~\ref{section:sensorExpansion}. 
Then, (3) ``Current features'' and (4) ``Your task'' are changed as below.

\subsubsection*{\textbf{(3) ``Current features'' :}} 
Here we explain the current computed features. 
At this point, explain using the words of position in the body.

\subsubsection*{\textbf{(4) ``Your task'' :}} 
Here we indicate what you want ChatGPT to output.
Since the purpose of this paper is to extend features, we specify that new features should be output.

\section{Experiment}
\label{section:experiment}
In this section, in order to evaluate the proposed prompts, we make a machine learning model to perform human activity recognition (HAR) and evaluate the accuracy and F1-score.

\subsection{Experiment setting}
\label{sectino:experimentSetting}

In the experiment, we use the Opportunity dataset~\cite{OPPortunityDataset1, OPPortunityDataset2}. 
This dataset includes data from 5 subjects, from accelerometers and IMU sensors attached to the full body. 
Additionally, this dataset has 5 classes of activity labels, which is ``Stand'', ``Sit'', ``Walk'', ``Lie'', and ``Others''. 
The machine learning model identifies these 5 activity classes from sensor data.



The experiment will be conducted using GPT-4 (May 24 Version), a feature of ChatGPT Plus. 
The input language is English.
In the sensor pioneering experiment, we use a new chat session which not have any chat history, with no history of the previous prompt. 
In the feature augmentation experiment, we use chat to continue the sensor pioneering experiment. 
In those experiments, 
we assume that the data collection and modeling were done using only elbow sensor data before inputting the prompt into ChatGPT. 

In the pre-processing, missing values were complemented with the average of the before and after values. 
For segmentation, we use the sliding window method with a window size of 5 seconds and an overlap rate of 30\%.



\subsection{Result}
\label{section:result}
In this section, we evaluate the results of the experiment and the accuracy of the activity recognition model created based on the experimental results. 
Figure ~\ref{fig:inputHAR1} - \ref{fig:outputHAR2} shows the input prompt which is described in section ~\ref{section:proposedMethod}, and the output result from ChatGPT. 

\subsubsection{Sensor Pioneering : } 
\begin{figure}[htb]
    \centering
    \includegraphics[width=.9\linewidth]{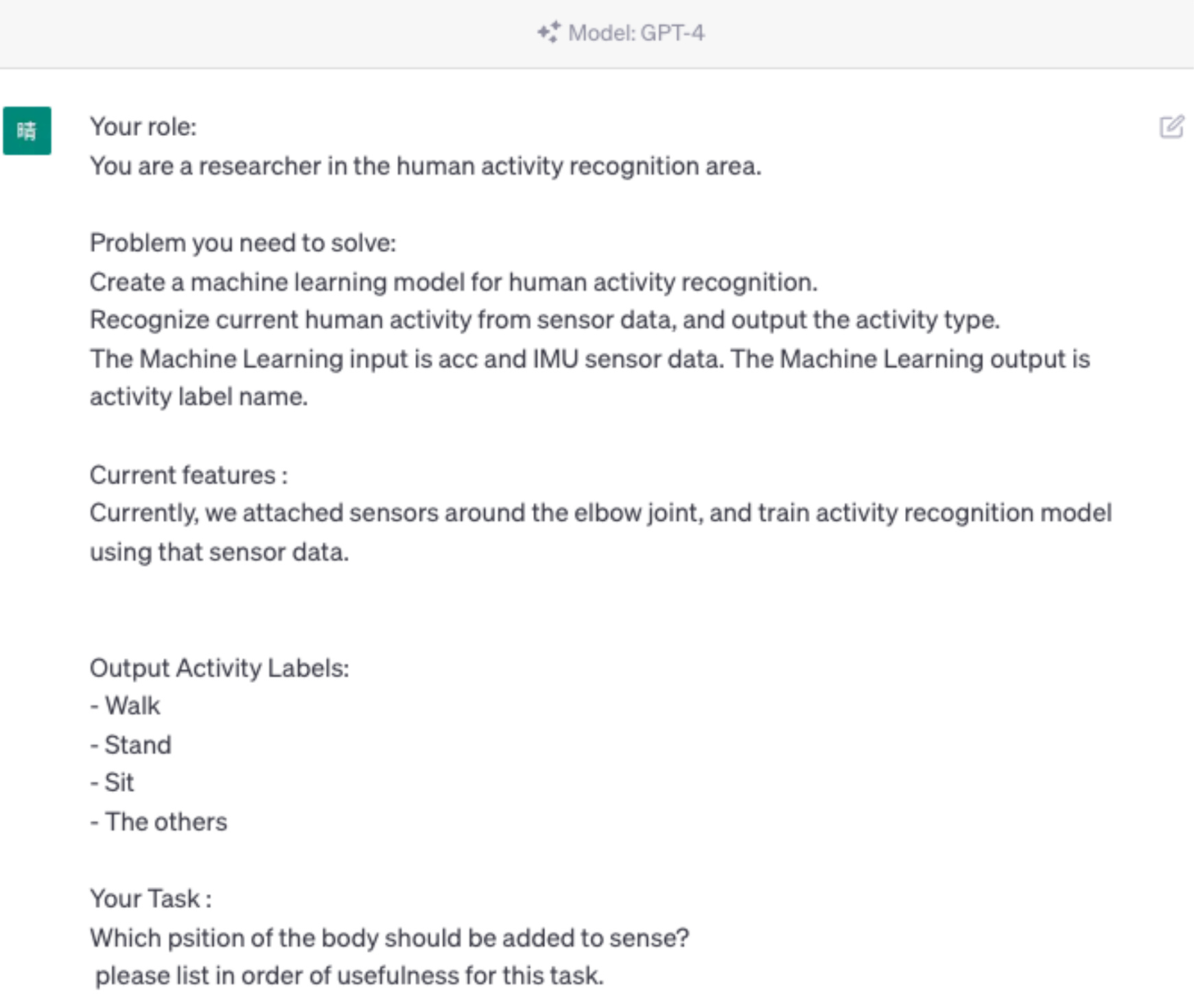}
    \caption{Input prompt for sensor pioneering in human activity recognition ( A: without (5) ``Current results'' ).}
    \label{fig:inputHAR1}
\end{figure}

\begin{figure}[htb]
    \centering
    \includegraphics[width=.9\linewidth]{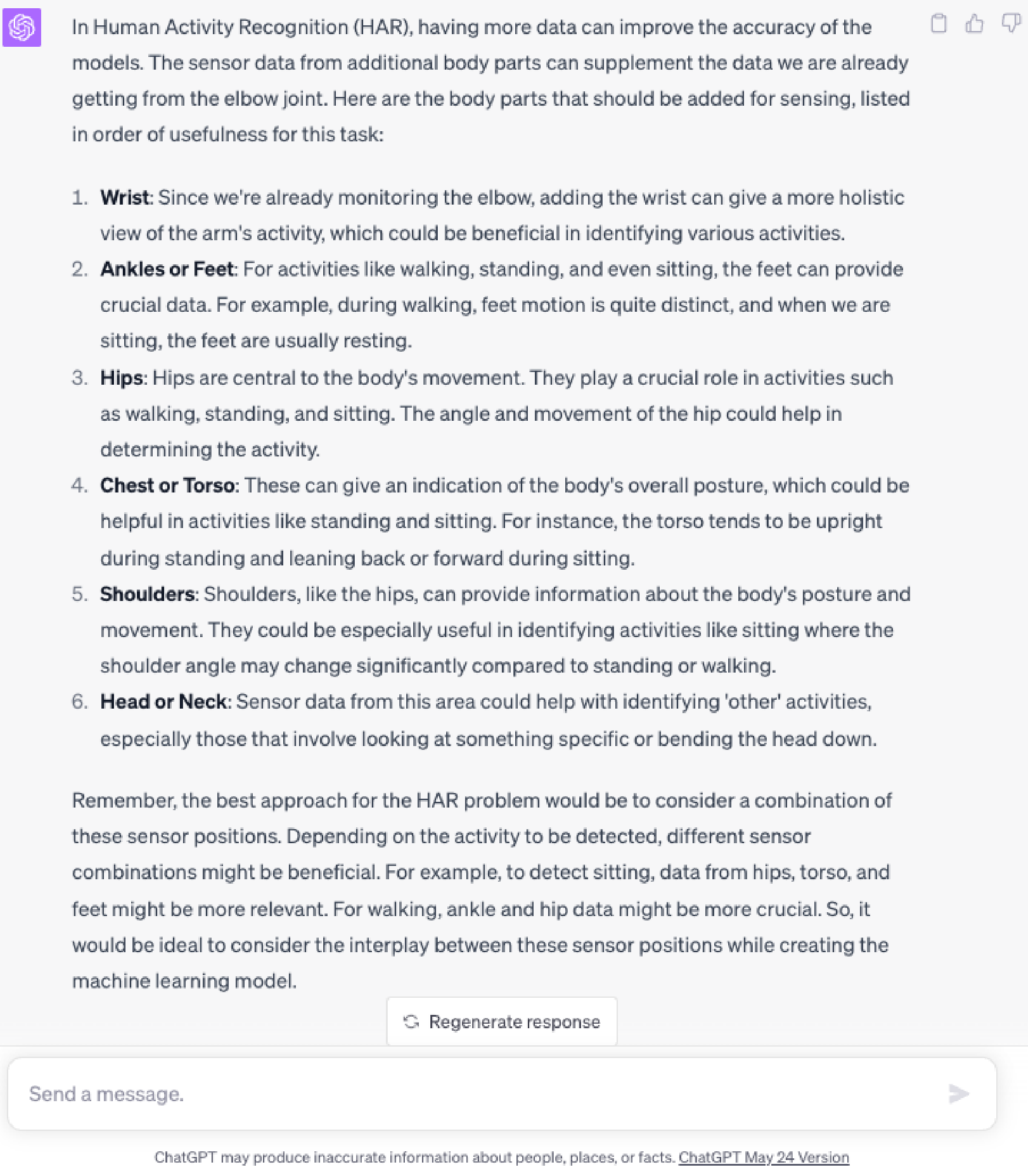}
    \caption{Output from ChatGPT for sensor pioneering in human activity recognition ( A: without (5) ``Current results'' ).}
    \label{fig:outputHAR1}
\end{figure}

\begin{figure}[htb]
    \centering
    \includegraphics[width=.9\linewidth]{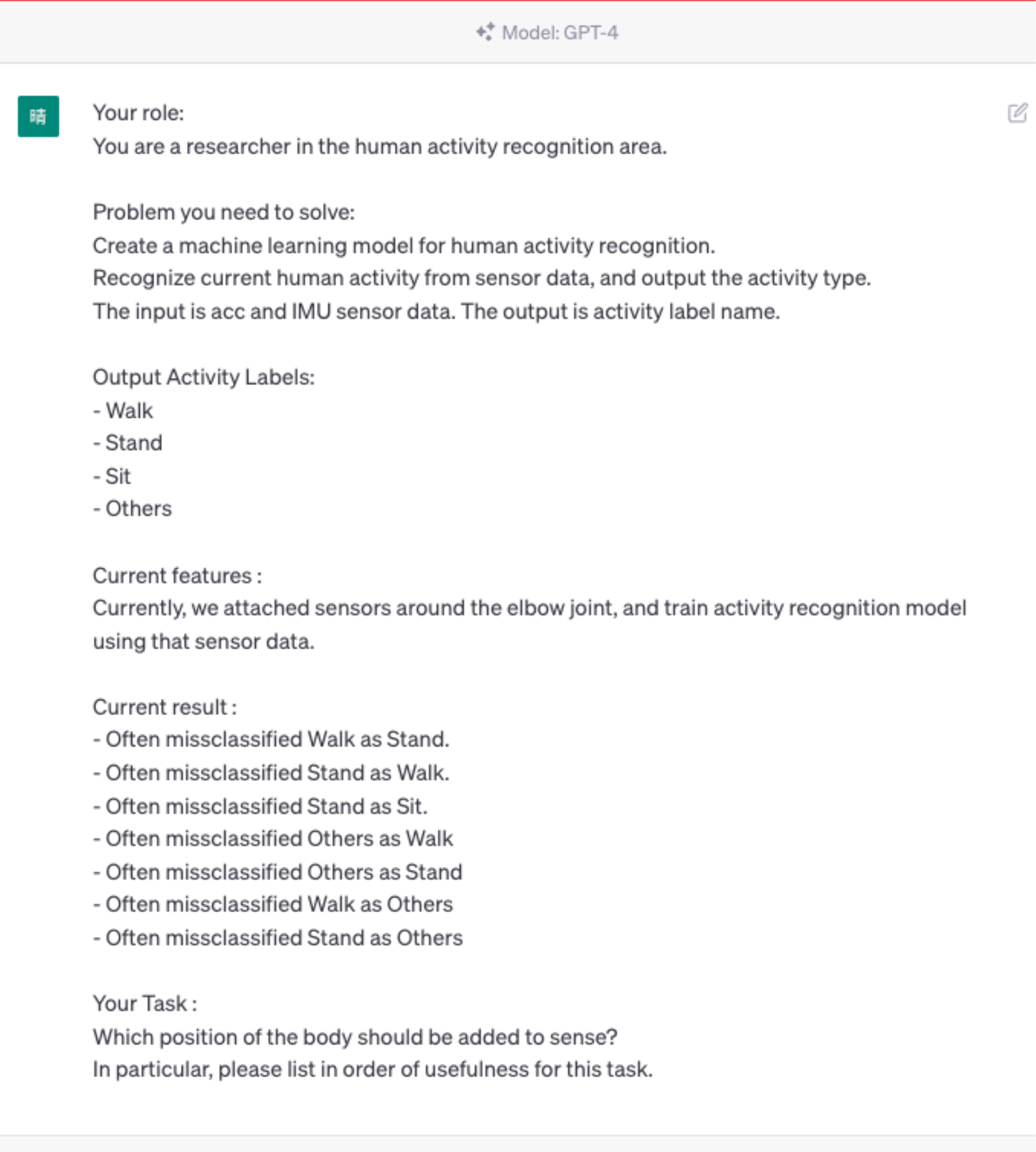}
    \caption{Input prompt for sensor pioneering in human activity recognition ( B: with (5) ``Current results'' ).}
    \label{fig:inputHAR2}
\end{figure}

\begin{figure}[htb]
    \centering
    \includegraphics[width=.9\linewidth]{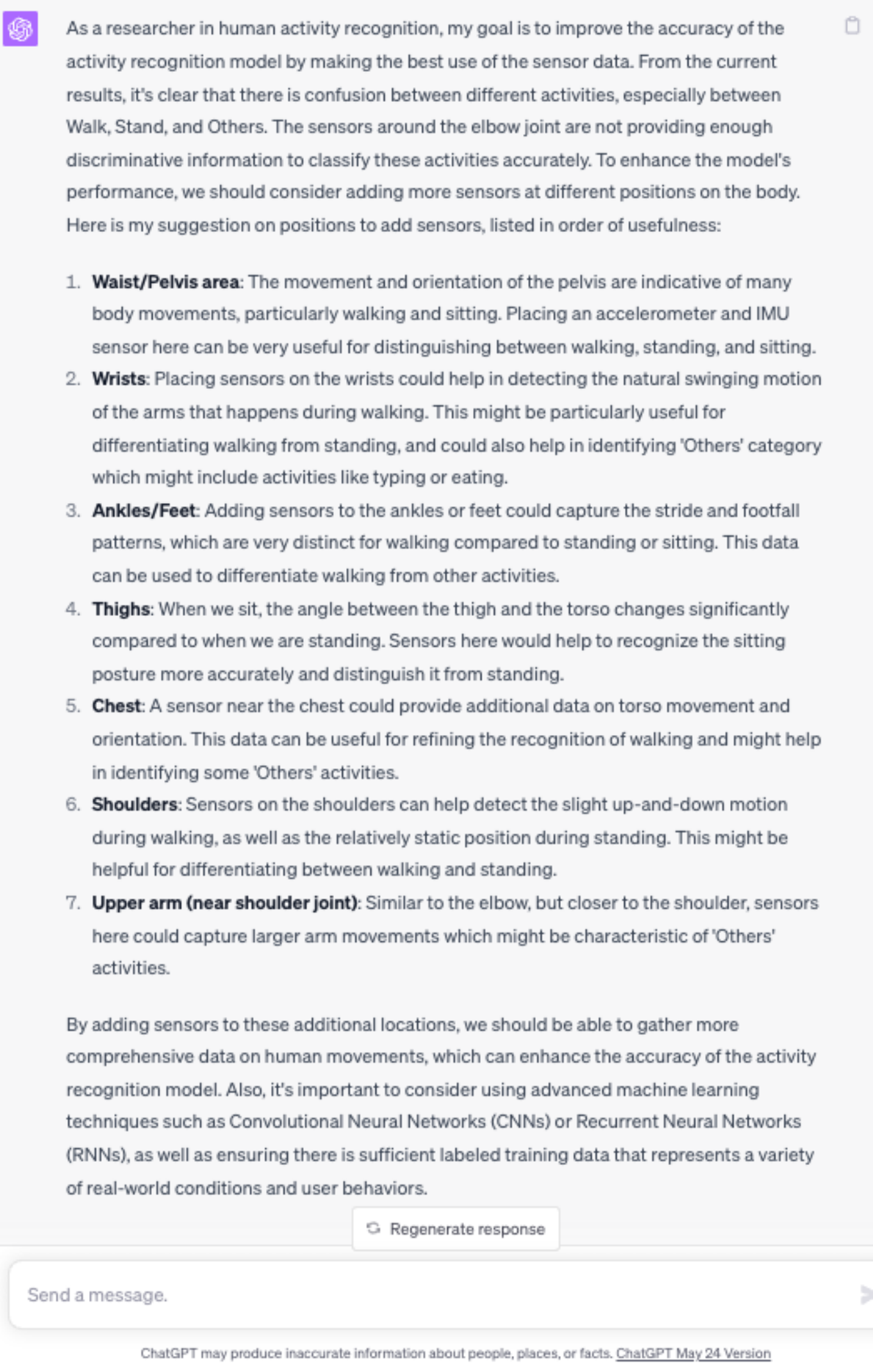}
    \caption{Output from ChatGPT for sensor pioneering in human activity recognition ( B: with (5) ``Current results'' ).}
    \label{fig:outputHAR2}
\end{figure}

In the experiment, we enter the 2 types of prompts, A) without (5) ``Current results'' and B) with (5) ``Current results''. 
Both of those promote entered into the initialized prompts. 
Figure ~\ref{fig:inputHAR1} and ~\ref{fig:outputHAR1} is the result of input without (5) ``Current results'', 
figure ~\ref{fig:inputHAR2} and ~\ref{fig:outputHAR2} is the result of input with (5) ``Current results''. 
In the baseline, we use the following 4 sensors ["RUA\^", "LUA\^", "RUA\_", "LUA\_"]. 

As shown in figure ~\ref{fig:inputHAR1} and ~\ref{fig:outputHAR1}, 
when we use prompt A) which does not include (5) ``Current results'', 
6 new sensor locations are suggested. 
Based on this, we decided to add the following 9 sensors ["R-SHOE", "L-SHOE", "RWR", "LWR", "RUA\_", "RUA\^", "LUA\_", "LUA\^", "HIP"].

As shown in figure ~\ref{fig:inputHAR2} and ~\ref{fig:outputHAR2}, 
when we use prompt B) which includes (5) ``Current results'', 
7 new sensor locations are suggested. 
Based on this, we decided to add those following 10 sensors ["R-SHOE", "L-SHOE", "RLA", "LLA", "BACK", "RUA", "LUA", "RWR", "LWR", "HIP"]. 
Comparing these figures, we can know that we can get more additional sensor suggestions when we input with  (5) ``Current results''.

\subsubsection{Feature Augmentation : } 
\begin{figure}[htb]
    \centering
    \includegraphics[width=.9\linewidth]{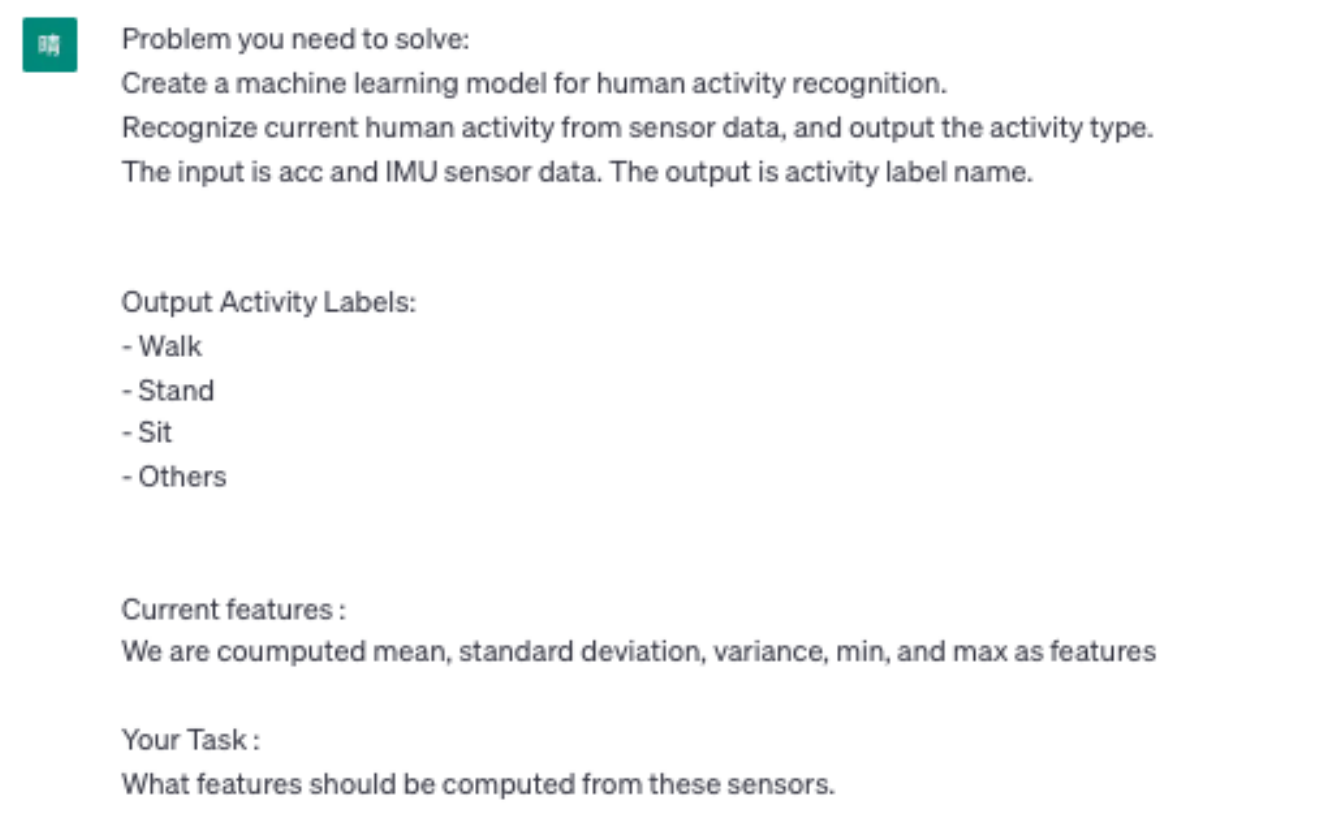}
    \caption{Input prompt for feature augmentation in human activity recognition.}
    \label{fig:inputFeatureCaluculation}
\end{figure}

\begin{figure}[htb]
    \centering
    \includegraphics[width=.9\linewidth]{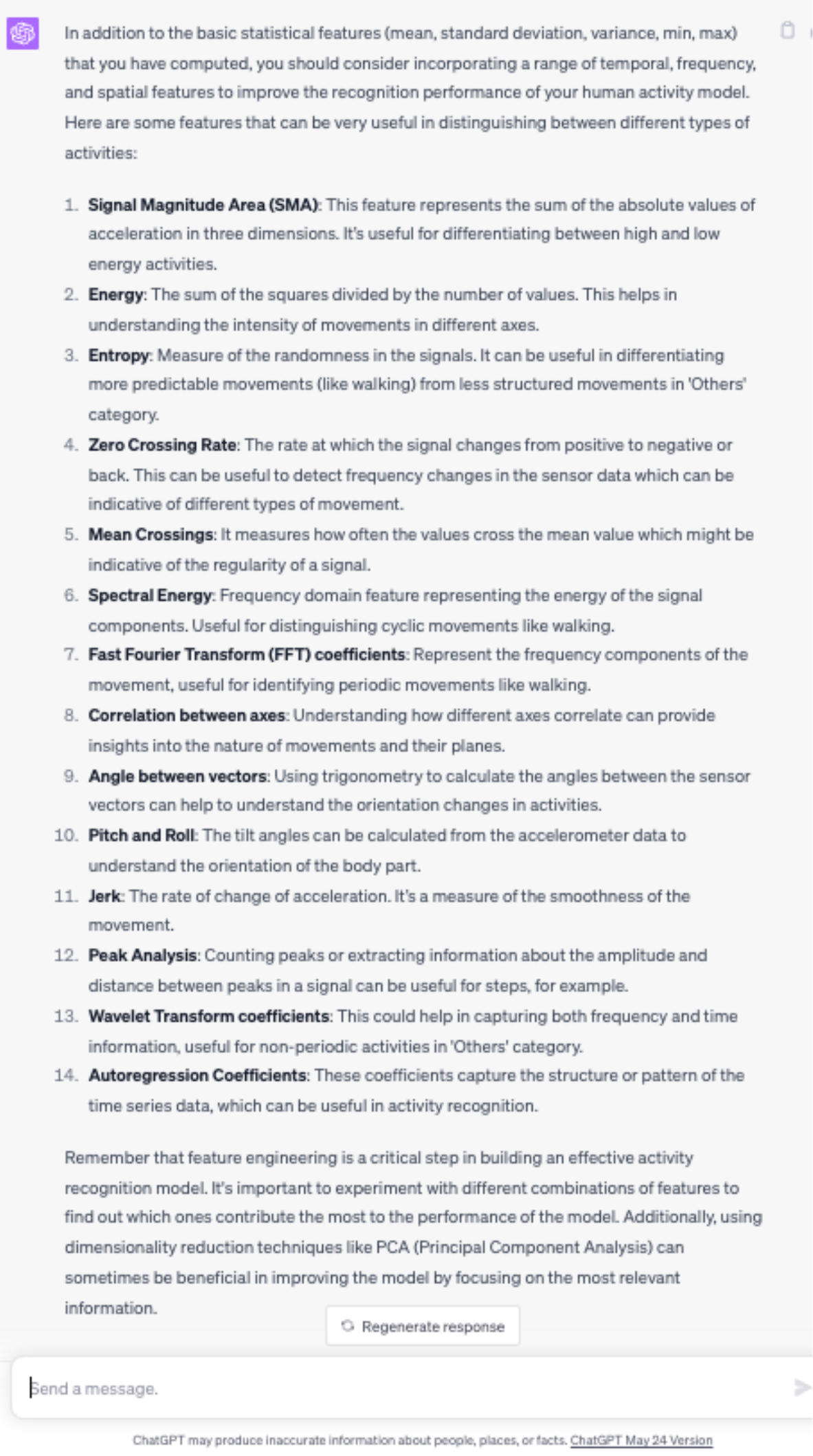}
    \caption{Output from ChatGPT for feature augmentation in human activity recognition.}
    \label{fig:outputFeatureCaluculation}
\end{figure}

Figure ~\ref{fig:inputFeatureCaluculation} - \ref{fig:outputFeatureCaluculation} shows the input prompt for feature augmentations proposed in section ~\ref{section:proposedMethod}, and the output result of ChatGPT. 
This result is the output when we input the feature augmentation prompt after the B) sensor pioneering prompt with (5) ``Current results''. 
In the baseline, we use the following 5 features ``mean'', ``standard deviation'', ``variance'', ``min'', and ``max''.

As shown in figure ~\ref{fig:outputFeatureCaluculation}, we got 14 types of features. 
Among the ChatGPT output, we added the following 10 features that are easy to calculate, 
``(1) Signal Magnitude Area (SMA)'', 
``(2) Energy'', 
``(3) Entropy'', 
``(4) Zero Crossing Rate'', 
``(5) Mean Crossing Rate'', 
``(7) Fast Fourier Transform (FFT) Coefficients'', 
``(8) Correlation between axes'', 
``(10) Pitch and Roll'',
``(11) Jerk'',
and 
``(12) Peak Frequency''
.

\subsubsection{Estimation Model :}
In this section, we evaluate the model using new sensors and features made by the proposed prompt. 
Table ~\ref{tab:result} are shown the Accuracy and F1-score. 
``Sensor Pioneering A'' means the prompts without (5) ``Current results'', and ``Sensor Pioneering B'' means the prompts with (5) ``Current results''. 

In ``(a) Baseline'', ``(b) All sensors'', and ``(d) Sensor Pioneering B'' the following features are used, ``mean'', ``standard deviation'', ``variance'', ``min'', and ``max'' which are the same when we describe the current features in the input prompt.

First, compared to ``(a) Baseline'',  ``(d) Sensor Pioneering A + Feature Augmentation'', and ``(f) Sensor Pioneering B + Feature Augmentation'' has higher accuracy. 
This is indicating that the proposed method is able to improve the accuracy.

Next, if we focus on evaluating the sensor pioneering method, ``(e) Sensor Pioneering B'' has higher accuracy than ``(a) Baseline''. 
On the other hand, ``(c) All sensors + Feature Augmentation'' has higher accuracy than ``(f) Sensor Pioneering B + Feature Augmentation''. 
However, the difference is not very large. 

Finlay if we focus on evaluating the feature augmentation method, 
``(f) Sensor Pioneering B + Feature Augmentation'' has higher accuracy than ``(e) Sensor Pioneering B''. 
This is the same in the comparison between ``(b) All sensors'' and ``(c) All sensors + Feature Augmentation''. 
This indicates that accuracy can be improved by adding features using the proposed method.

\begin{table}[htb]
    \centering
    \caption{Accuracy\% / F1-score\% comparison between the model using features calculated by the proposed method and previous studies.}
    \begin{tabular}{l c}
    \hline


    Model                                    & Average    \\ \hline
    (a) Baseline                                 & 74.3\% / 75.4\% \\ \hline
    (b) All sensors                              & 81.2\% / 82.9\% \\ \hline
    (c) All sensors + Feature Augmentation        & 83.3\% / 84.8\% \\ \hline
    (d) Sensor Pioneering A + Featrue Augmentation & 78.5\% / 80.6\% \\ \hline
    (e) Sensor Pioneering B                       & 80.5\% / 82.0\% \\ \hline
    (f) Sensor Pioneering B + Featrue Augmentation & 81.3\% / 82.8\% \\ \hline
    
    \end{tabular}
    \label{tab:result}
\end{table}

\section{Discussion}
\label{section:discussion}

We obtain the accuracy and F1-score of the All sensors model and the model made by the proposed prompt are not significantly different in table ~\ref{tab:result}. 
This is also the same in both of the comparisons by subject or by average. 
In addition, while the ``(b) All sensors'' mode uses 18 sensors, the ``(e) Sensor Pioneering B'' only uses 11 sensors. 
This means that the proposed prompt is possible to efficiently find important sensor locations and features.

In the feature augmentation, 
``(f) Sensor Pioneering B + Feature Augmentation'' has higher accuracy than ``(e) Sensor Pioneering B''. 
In other words, the overall accuracy is better when features are added using the proposed method. 
However, the improvement is very small, and hardly any changes at all. 
In the experiment, we use initialized prompts with no previous conversation history when we input the prompt. 
However, as the experiment progressed, it became changed that ChatGPT was showing more and more suggestions which is new sensor locations and new feature augmentations. 
During this, all experiments were conducted using ChatGPT May 24 Version, and we do not know the reason why the number of suggestions increased. 
However, one possibility is that my other chat history was used as background information for input in ChatGPT.

\section{Conclusion}
\label{section:conclusion}
In this paper, we propose a feature engineering method using Large Language Models (LLMs) in order to efficient sensing and feature selection. 
As a result, we got new sensor locations from ChatGPT. 
Then, we make a machine learning model by using the proposed method. 
Although the number of sensors in the model made by the proposed method is small, the model shows almost the same accuracy as the mode by the all of sensors in the Opportunity dataset. 
This indicates that the proposed method is able to extract important features efficiently. 
In addition, in the augmentation of the features, we obtained new feature calculations, 
and show that the prediction accuracy could be improved over the existing method.

Our proposed method can be used to know the critical sensing location before sensing. 
However, when looking at the output of ChatGPT, the output describes features that are important in general human activity recognition. 
Therefore, in order to support special types of behavior, it may be necessary to devise a way to input activity labels name with more detailed descriptions instead of label names. 


\bibliographystyle{ACM-Reference-Format}
\bibliography{mybib}

\end{document}